\documentclass[a4paper,12pt]{article}
\usepackage[a4paper, margin=0.70in]{geometry}
\usepackage[utf8]{inputenc}
\usepackage[T1]{fontenc}
\usepackage{amsmath, amssymb}
\usepackage{geometry}
\usepackage{graphicx}
\usepackage{array}
\usepackage{hyperref}
\usepackage{soul}
\newcommand{\com}[1]{} 
\newcommand{\uncom}[1]{#1} 

\usepackage{tikz}
\usetikzlibrary{calc, shapes, arrows}
\usepackage{draftwatermark} 
\SetWatermarkText{}
\SetWatermarkScale{5}
\SetWatermarkAngle{45}
\SetWatermarkColor{black!0} 
\usepackage{amsfonts}
\usepackage{url}
\usepackage{textcomp}
\usepackage{subcaption}
\usepackage[numbers]{natbib}
\title{Monetary Macro Accounting Theory}
\author{Dr.\,Ren\'ee Men\'endez, Dr.\,Viktor Winschel\footnote{viktor.winschel@gmail.com, mobile +49 179 7621055}}
\date{\today}
\begin{document}
\maketitle
\setcounter{tocdepth}{2}
\begin{center}
\begin{minipage}{\textwidth} 
\uncom{
{\scriptsize
\tableofcontents}
\begin{abstract} \footnotesize 
We develop a monetary macro accounting theory (MoMaT) and its software specification for a consistent national accounting. 
In our money theory money functions primarily as a medium of payment for obligations and debts, not as a medium of exchange, 
originating from the temporal misalignment where producers pay suppliers before receiving revenue. 
MoMaT applies the legal principles of Separation and Abstraction to model debt, contracts, property rights, and money to understand their nature. 
Monetary systems according to our approach operate at three interconnected levels: 
micro (division of labor), meso (banking for risk-sharing), and macro (GDP sharing, money issuance). 
Critical to money theory are macro debt relations, hence the model focuses not on the circulation of money but on debt vortices: 
the ongoing creation and resolution of financial obligations.
The Bill of Exchange (BoE) acts as a unifying contractual instrument, linking debt processes and monetary issuance across fiat and gold-based systems. 
A multi-level BoE framework enables liquidity exchange, investments, and endorsements, designed for potential implementation in blockchain smart contracts 
and AI automation to improve borrowing transparency. 
Mathematical rigor can be ensured through category theory and sheaf theory for invariances between economic levels and homology theory for monetary policy foundations.
Open Games can structure macroeconomic analysis with multi-agent models, making MoMaT applicable to blockchain economic theory, 
monetary policy, and supply chain finance.
\end{abstract}
}
\end{minipage}
\end{center}
\newpage
\section{Introduction}\label{sec:intr}
Economists since Adam Smith in 1776 in \emph{The Wealth of Nations}~\cite{smith_wealth_1776}
ask the question: \emph{What is money?}
Martin Hellwig in \emph{The Challenge of Monetary Theory}~\cite{Hellwig1993} remarks that we have no conceptual foundations for monetary economics 
and doubts that money is a phenomenon to be discussed within Walrasian market theory.

Since then new perspectives on the nature of money have been provided.
Money emerges in search theoretic models~\cite{kiyotaki1993search} as a medium of exchange. 
\emph{Money is Memory}~\cite{kocherlakota1998money} relates to imperfect record-keeping and frictions like limited enforcement and anonymity. 
Heterodox approaches like \emph{Modern Money Theory}~\cite{wray2012modern}, ~\cite{kelton2020deficit} 
emphasise the role of states and central banks in defining and managing money beyond a market framework. 
Walrasian approaches of Dynamic Stochastic General Equilibrium~\cite{gertler2011financial} incorporate money 
as \emph{cash-in-advance} constraints but struggle to account for financial crises. 
Cryptocurrencies have revived debates on the definition of money, revisiting record-keeping challenges extending \emph{Money is Memory}. 
Accordingly, we will relate our approach to the concepts in blockchain research.
Institutions like the Bank of England~\cite{mcleay2014money} contributed practical insights into money issuing and
continue to explore central bank digital currencies (CBDCs)
and keep on asking the question: \emph{What is money?}~\cite{starr2016responding}

We develop a monetary macro accounting (MoMa) theory (MoMaT) to function to facilitate division of labor
in decision on supply chains:\footnote{
See~\cite{menendez2010theorie} for a precursor of a money theory proposed in this paper.}
\emph{Money is needed because production takes time, requiring producers to incur debts to pay suppliers and employees 
before receiving payments from customers.}
Money serves as a \emph{medium of payment} and \emph{medium of exchange}
used in purchase or loan obligation and disposition contracts on transferring property rights on money or 
products in economies characterised by division of labor.

MoMaT does not justify relative prices or assign intrinsic value to money. 
It is based on contractually recorded value judgments made by economic agents. 
Money functions to record and settle debts arising from contracts but does not itself require having some inherent value for its role in obligation settlements.
In order to understand money in debt relations,
we look at exchanges involving bookings in two double-entry accounting systems,
often termed \emph{quadruple accounting}.\footnote{See~\cite{sna_website_2008} for a quadruple \emph{System of National Accounts 2008}.} 
We use the term \emph{macro accounting} distinguishing it from traditional \emph{micro or double-entry accounting}. 
Beyond the micro level of the division of labor and debt networks, 
we discuss the meso level where the risk of failing investments is distributed by banks across a portfolio of investments.
At the macro level the GDP is shared, risk is socialised, and money is created, issued, and circulated by central banks.
We use the legal \emph{principles of Separation and Abstraction} to clarify conceptual challenges in monetary theory, 
particularly concerning the nature of debt and payments answering what money is by typing money in obligation relations 
as the fulfillment thing for disposition contracts on property rights between products and money.

MoMaT uses the Bill of Exchange (BoE) instrument as principles 
to understand paper and gold-based money over the hierarchical levels of economies.
A BoE is a formal financial version of the informal \emph{I owe you} contract of a bar beer coaster.
We connect to sheaf theory as the mathematics of the transition from local to global structures 
to model the accounting invariances of Assets, Equity and Liability at the micro level.
The meso and macro are invariances of debt and payment relations as sums of receivables and liabilities 
and payment relations as sums of expenditures and revenues.
Open games can implement macro accounting systems, smart contracts and structural econometric models to analyse macro accounting data.
MoMaT applies to monetary policy, supply chain finance, or platform-based demand-supply matching.
\section{Microeconomics}\label{sec:micro}
Macro accounting records networks of debts in economies characterised by division of labor.  
To understand them, we use the legal Principles of Separation and Abstraction from the German Civil Code (\emph{Bürgerliches Gesetzbuch}, BGB).  
They delineate fundamental concepts like claims, receivables, debts, liabilities, money or payments.

\begin{figure}[t!]
    \centering
    \includegraphics[width=0.57\textwidth]{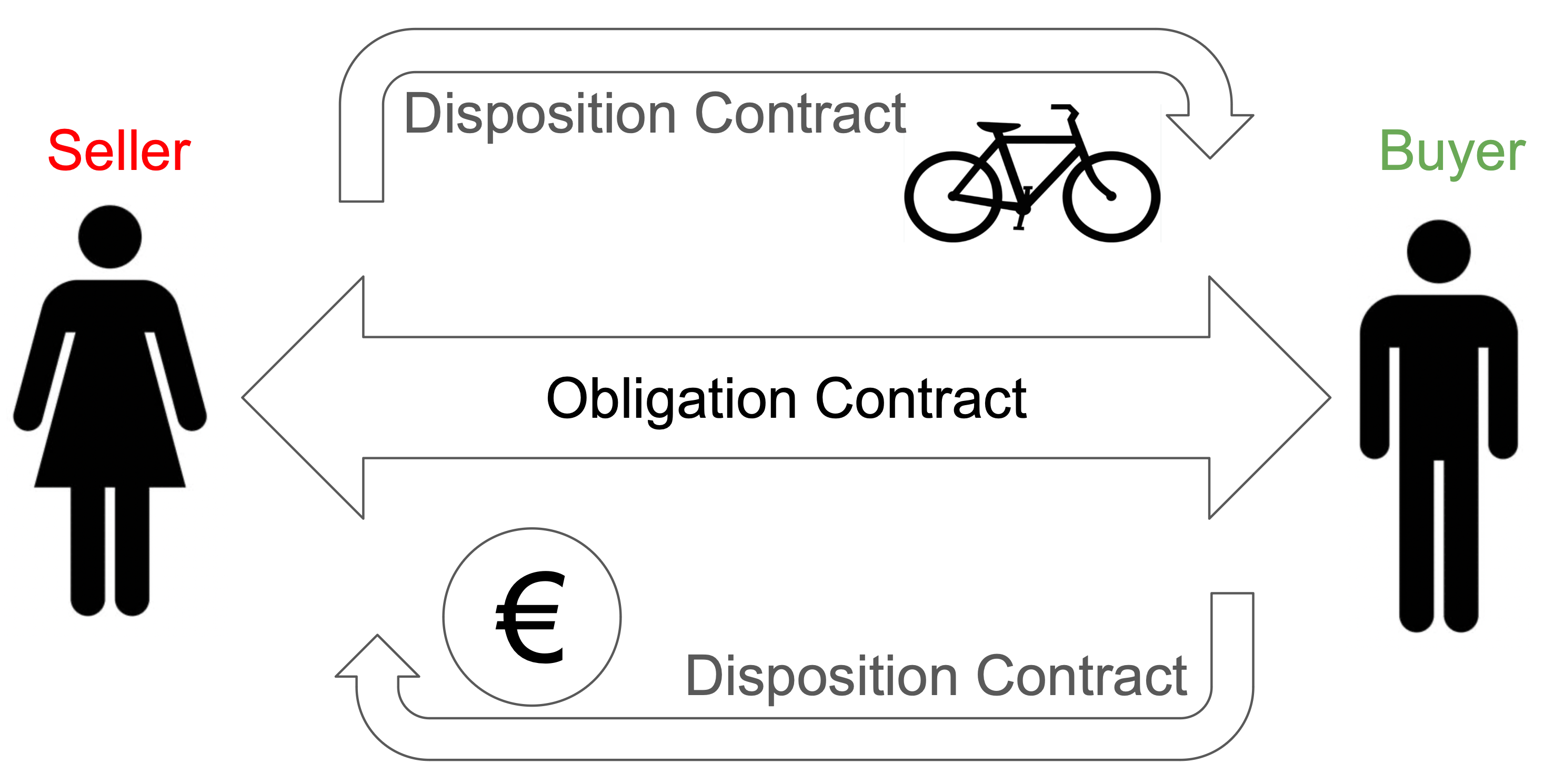}
    \caption{Obligation and Disposition Contracts}
    \label{fig:obdis}
\end{figure}

The {\bf Separation Principle} asserts that the obligation contract which establishes a debt relation
and the disposition contracts which govern the transfer of property rights are associated but distinct legal transactions.  
Obligation contracts involve two parties, like in purchase, work or loan contracts, illustrated in Figure~\ref{fig:obdis}.
In exchanges of goods, money, or loans, the structure consists of
the {\bf obligation} contract that governs the {\bf agreement} to exchange, establish, modify, or extinguish rights, like ownership of a bicycle and money,
and two {\bf disposition} contracts to {\bf fulfil} the obligation, like to transfer ownerships on the exchanged.

Payments are not always executed through the physical handover of money. 
Instead, another contract on bank deposits enables payments by transferring money ownership to the payee.  
This involves a crucial distinction: bank deposits are not money, but option rights
allowing customers to instruct their bank to withdraw money or to transfer the deposit to others.
The role of a bank in the deposit obligation contract is to await customers' instructions before fulfilling the disposal contract by executing a money transfer or withdrawal. 
We denote transfer rights over money in deposits as "money" or commands for money.

The {\bf Abstraction Principle} states that the {\bf validity of the obligation} contract is {\bf independent of the validity of the disposition} contracts.  
Even if an obligation contract is redeemed invalid, ownership transfers executed through disposition contracts remain legally effective. 
Likewise, if a disposition contract is invalid, the obligation contract remains unaffected.
This can create imbalances, where one party acquires ownership without a valid contractual basis. 
To rectify such situations § 812 BGB provides a restitution claim
and ensures that the party who benefited without a legal basis must return the asset or compensate the counterparty.  

In the principles applied to purchase a product for money (§ 433 BGB),
the obligation contract commits the seller to deliver a product and the buyer to provide payment.
The disposition contracts transfers the ownership (§ 929 BGB) of the product from seller to buyer and money from buyer to seller.

In the principles applied to a loan contract (§ 488 BGB),
the obligation contract is about a loan that obligates the bank to grant access to money and the debtor to repay it.
The two disposition contracts are:
the bank grants a deposit, enabling the debtor to command central bank money transfers or withdrawals
and the debtor repays the loan, settling the obligation.

Misconceptions in monetary theory often arise from a failure to apply these principles correctly.  
An example is the concept of \emph{Kreditgeld} (German for \emph{loan money}). 
Here, a contradiction emerges:  A loan is a two-sided obligation contract, establishing mutual commitments between debtor and creditor.  
Money, on the other hand, serves as a means to fulfill an obligation, facilitating settlement rather than creating new contractual obligations.  
By conflating loans and money, such interpretations blur the distinction between the creation of debt and the fulfillment of debt, 
leading to conceptual inconsistencies in monetary theory.
\subsection{Supply Chains and Pins}\label{sec:scap}
\begin{figure}[htbp]
    \centering
    \includegraphics[width=0.8\textwidth]{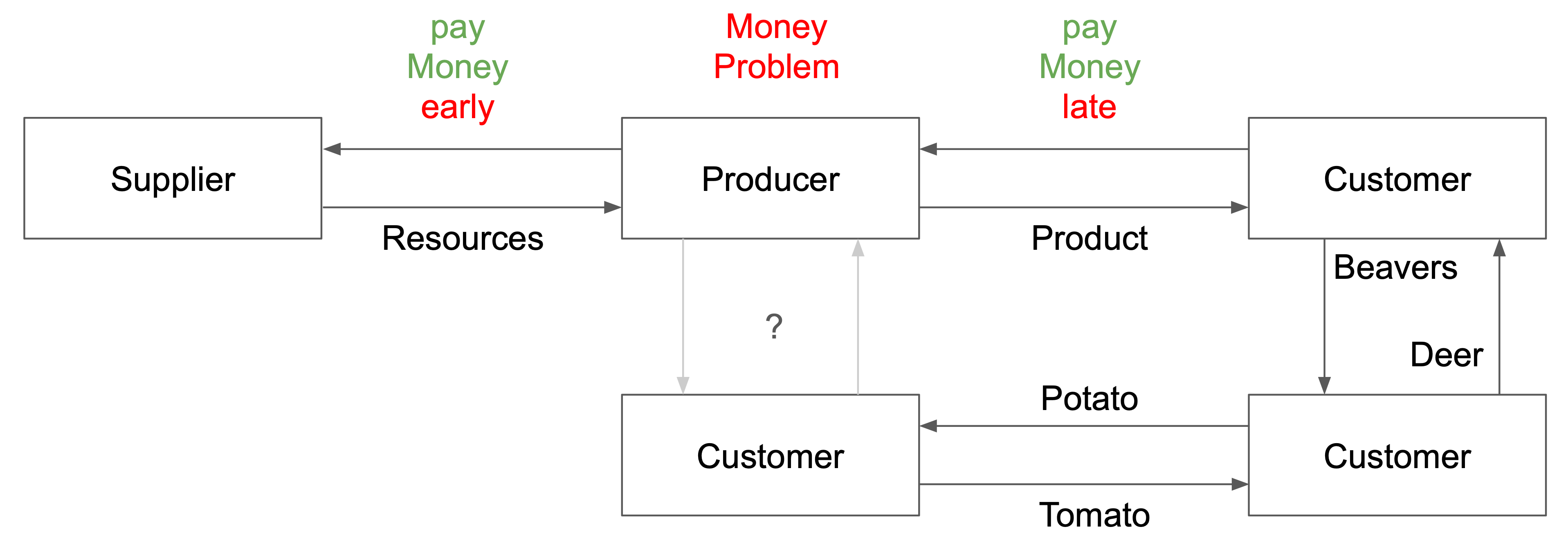} 
    \caption{Money Demand in Supply Chains}
    \label{fig:liquidity}
\end{figure}

Figure~\ref{fig:liquidity} shows the need for integrating obligations in supply chains of economies organised by a division of labor.
Money exists because producers need investments to pay suppliers of resources before being paid by customers.
The double coincidence of wants story with \emph{money as a medium of exchange} in time
misses to capture this functionality needed to intermediate between producers and customers over time.

A similar pattern is in the bookings of a Bills of Exchange (BoE)~\cite{radke2019buchfuhrung}:
\emph{When buying goods, two different interests meet. 
The supplier of inputs of production wants to receive his money as early as possible, 
since he has made payments in advance in various ways 
- he has produced, processed or procured the goods at his own expense. 
The producer wants to pay as late as possible, 
as he still has to sell the goods and only receives the money from his customers later}
(own translation, \emph{buyer} replaced by \emph{producer}).

The same pattern appears in the first sentence of \emph{The Wealth of Nations}~\cite{smith_wealth_1776}, 
followed by the famous example of the production of pins:
	\emph{The greatest improvement in the productive powers of labor, and the greater part of the skill, dexterity, and judgment with which it is
	any where directed, or applied, seem to have been the effects of the
	division of labor.{[}...{]} One man draws out the wire, another
	straights it, a third cuts it, a fourth points it, a fifth grinds it at the top for receiving the head [...]}.

Economies with a division of labor are sequences of decentralised resources transformations into products 
spread over production steps, producers, and consumers.
A producer is first indebted and second does not only (if at all) consume his product but also other goods.
He sells his product for money, which allows him to buy other goods and to repay debts to investors to pay workers and resource suppliers. 
Workers are not indebted from the beginning; they use money for consumption or saving.

To emphasise the distinction between production and consumption, we use two terms for products:  
Commodities serve as a means to attract money to producers.
Goods provide utility through consumption to consumers.
Although it may appear that products and money circulate in cycles, 
their movement are rather vortices in time, connecting past and future obligations. 
While money circulates as paper, the paper itself remains the same, facilitating different obligation and disposition contracts. 
The key focus is on debt contracts and how they originate, are cleared, and eventually conclude. 
Money flows through different transactions, repaying distinct debts in different contractual contexts.

BoEs have a finite lifespan, ending at maturity, once the BoE is settled, causing the debt relation and the need for money to cease. 
This contrasts with the concept of money circulation; debt relations continuously emerge and disappear 
which form the entities of interest in a money theory.

In Figure~\ref{fig:liquidity} the question mark denotes that
theories of money as a \emph{medium of exchange} overlook the crucial distinction between commodities and goods.
The opposing demands for products and money between producers and consumers naturally align  
and eliminate the need for a value theory of money in sequential exchanges of consumers.
A value argument is needed to argue why to accept a worthless money for one's own valuable resources in order to exchange afterwards 
for another valuable consumption good.
The key insight is that producers demand money to repay debts, while consumers acquire goods for money. 
Since these flows match, monetary theory can be based on monetary macro accounting (MoMa) rather than subjective valuations of money.
\subsection{Prices and Beavers and Deer}\label{sec:pbad}
Figure~\ref{fig:liquidity} also shows another starting point of monetary theories: 
some products are to be exchanged, here beavers and deer, as in~\emph{The Wealth of Nations}~\cite{smith_wealth_1776}, in chapter 6: 
\emph{If among a nation of hunters, for example, it usually costs
twice the labor to kill a beaver which it does to kill a deer, one
beaver should naturally exchange for or be worth two deer. It is natural
that what is usually the produce of two day's or two hours labor, 
should be worth double of what is usually the produce of one day's or one hour's labor.}

Without money, the number of beavers exchanged for the number of deer represents their relative value.
The \emph{natural} theory argues that the relative price should depend on the relative amount of labor used to produce the products. 
The cause of the existence of money is then that money enables exchanges even if owners of deer do not want to exchange them for beavers, tomatoes or potatoes.
This overly restricted requirement is called the \emph{double coincidence of wants}. 
Money relieves it and enables otherwise impossible exchanges.
Money is conceptualised as a num\'eraire or neutral commodity for exchange.
The value of money is then for example given by the welfare increase from enabling exchanges.

In MoMaT the story can be told by debt relations rather then by money value and relative prices.
If owner B of beavers b, measured in units of b, denoted by [b], and the owner D of deer d in [d] want to exchange 1 [b] for 1 [d],
they record the difference in value, at an agreed upon price of say 2 [b] for 1 [d], i.e. 2/1 [b/d].
This might have been the birth of the marks, measured in [b], on tally sticks, documenting the claim of D to B.
If D has trained his son in hunting d at no costs for D, he can be taken as a child abusing entrepreneur, who made a profit from a markup over his costs.
The next day D gets one [b] from B, the debt relation ceases to exist, and the tally stick can be thrown away, used as a tooth brush 
or for burning down the British parliament, as in 1876~\cite{tally_stick}, when wooden tally sticks where abandoned 
after they have been used for 40,000 years as a debt recording instrument.

In a gold-based system prices are expressed in gold [kg of gold/b] and converted into money prices [\$/b] via money as a num\'eraire [kg of gold/\$].
In fiat systems prices are directly measured in monetary units [\$/b] without requiring a num\'eraire.
The specific currency like Dollar or CryptoToken is irrelevant; 
what matters is that it is a legally defined means of payment, enabling the settlement of debts from obligation contracts through disposal contracts.

Money does not need intrinsic value for producers either. 
They calculate input and output prices in monetary units, with profits as a markup, where the monetary unit cancels out. 
Intrinsic value would be counterproductive, obscuring this ratio. 

Assigning value to money resembles a Russell paradox (of sets containing themselves), suggesting a need for type theory in monetary theory. 
A higher-order type theory could model the creation of currencies as the type of a type-generating operation: 
e.g. where the Euro and the ECB are created rather than issued Euros by the ECB.
A Russell paradox of central banking is analogous to: \emph{Who shaves the barber who shaves all those who can not shave themselves?}, 
as: \emph{Who provides the means of payments to the bank 
which provides the means of payments to those banks which can not provide the means of payments to themselves?}

Rather than a value theory of money we need money to represent:
(A) what is to be paid at the beginning, 100.000~[\$], and repaid at the end of an investment, 110.000~[\$], 
with the interest rate being 10.0\%=110/100 - 1.0,
(B) a price of a product, 6.59~[\$/BigMac],
(C) a relative price of two products, 2.45~[BigMac/Cheeseburger] $\approx$ 6.59~[\$/BigMac]~/~2.69~[\$/Cheeseburger],
and (D) a ratio of costs to output prices for calculating profits by markup pricing, 
like in $\pi=p-c=mc-c$ where $\pi$ := profit~[\$/product], $p$ := price~[\$/product], $m$ := markup~[$\mathbb{R}, >1.0$], $c$ := costs~[\$/product].
\section{Mesoeconomics}\label{sec:meso}
At the meso level banks manage loans, investments, repayments, clearing, and risk allocation. 
Failed investments are to be booked as consumption expenditures, reducing investor's wealth. 
Risk absorption involves three layers: 
producers absorb failures of unproductive use of money. 
Banks distribute risks via interest rates over portfolios of failed and successful investments into producers.
Central banks act as unlimited risk absorbers of failed banks.

To properly account for returns and losses from investment, it is crucial to note that banks do not create money but only references to money in deposits. 
They can instruct the central bank to issue cash but must first incur debt at the central bank to generate loans from the central bank, as well.
Essentially, banks produce solvency for companies using money as input, with the supply chains being central bank, bank, company.

Modern liberal societies, structured around obligation and disposition contracts for property transfers, operate as follows:  
1. Central banks issue means of payment to settle monetary obligations.  
2. Banks finance investments using central bank money.  
3. Companies use loans to pay resources, labor and insurance premiums (interest rates) for the possibility to not being able to repay the loans.
4. Consumers or households consume products produced from the resources and companies they own.  
\subsection{Bill of Exchange}\label{sec:boe}
A Bill of Exchange (BoE) is a financial instrument used in trade and commerce to provide credit and facilitate payments. 
It functions as an unconditional disposition contract within an obligation contract, where the {\bf drawer} issues the BoE, 
ordering the {\bf drawee} to pay a specified amount to the {\bf payee} often the drawer at a future date or on demand, at a discount or interest rate. 
The drawee accepts the order by signing it, becoming the {\bf acceptor} commited to pay.

\begin{figure}[htbp]
    \centering
    \begin{subfigure}[b]{1.0\textwidth}
        \centering
        \includegraphics[width=0.8\textwidth]{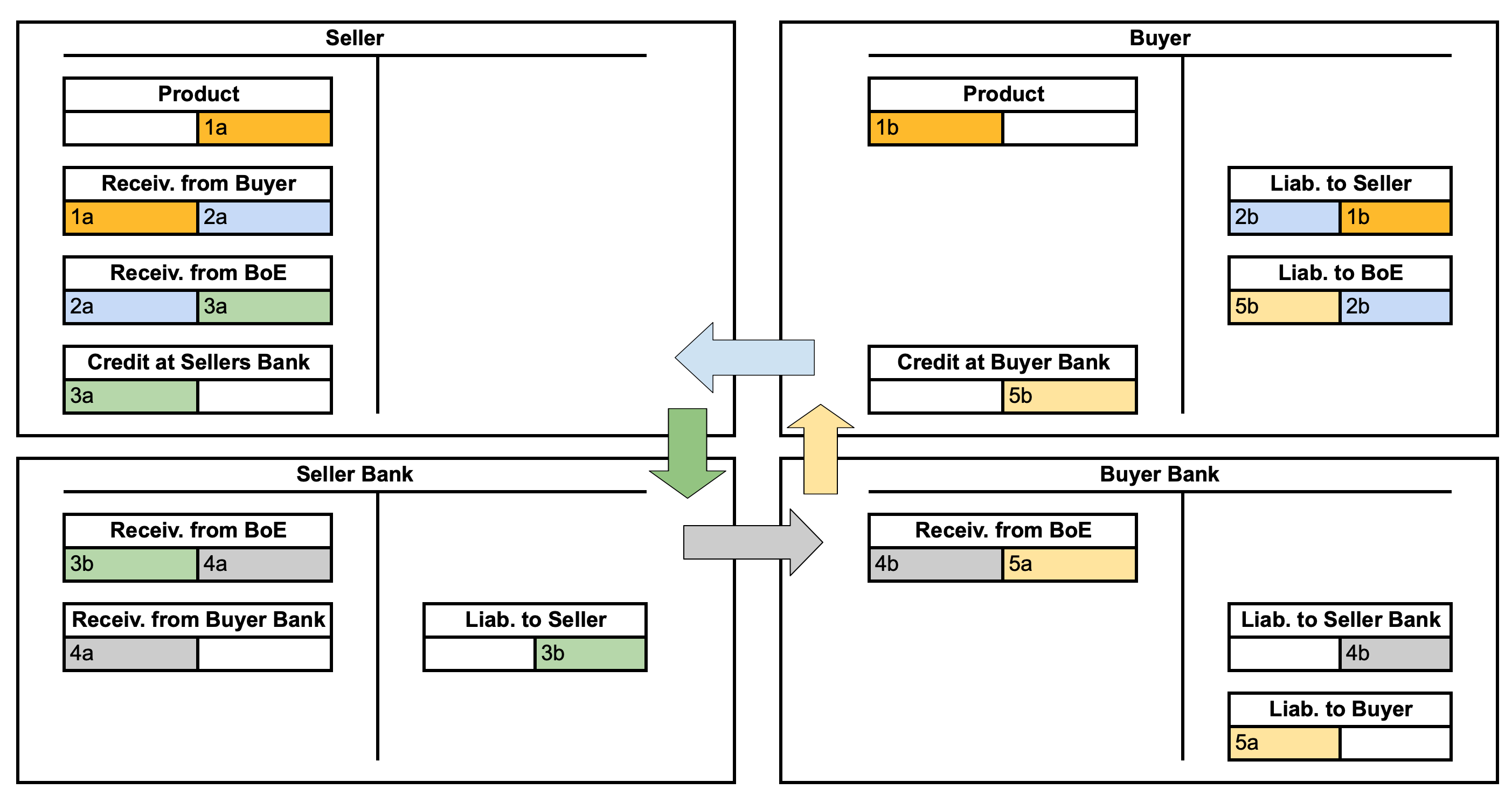}
	\caption{Bookings of a Bill of Exchange}
	\label{fig:boe_vor}
    \end{subfigure}
    \hfill \vspace{0.1cm}
    \begin{subfigure}[b]{1.0\textwidth}
        \centering
        \includegraphics[width=0.8\textwidth]{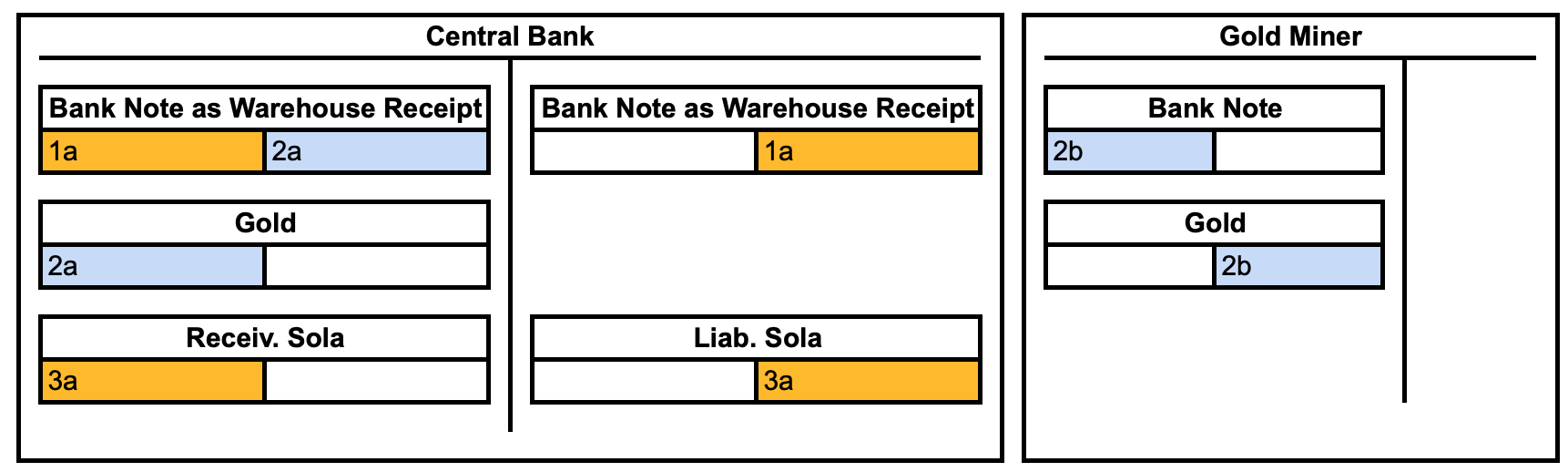}
        \caption{Issuing Bank Notes in Gold-Based Monetary Systems}
        \label{fig:gold}
    \end{subfigure}
    \hfill \vspace{0.1cm}
    \begin{subfigure}[b]{1.0\textwidth}
        \centering
        \includegraphics[width=\textwidth]{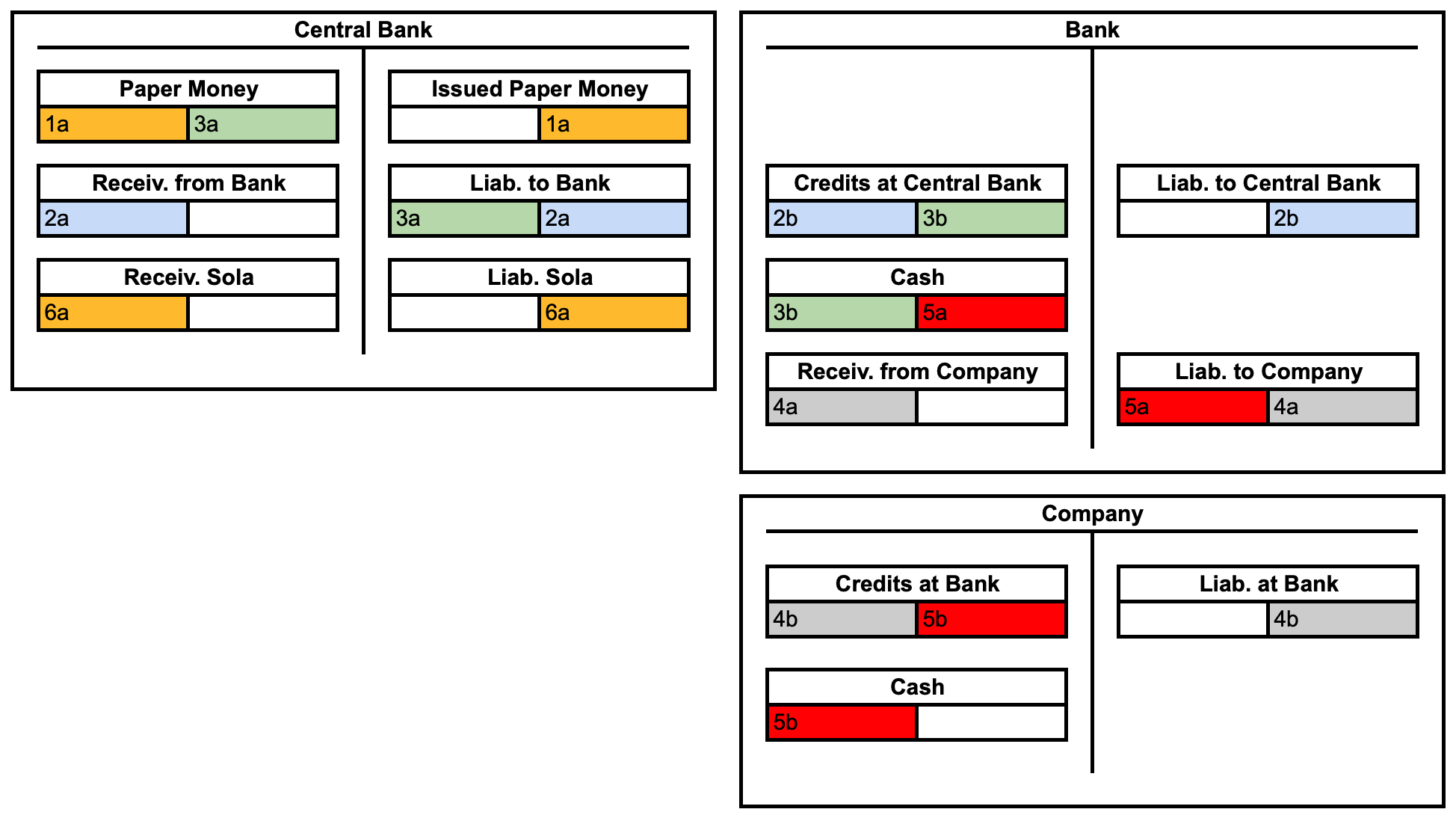}
	\caption{Issuing Paper Money in Fiat Monetary Systems}
        \label{fig:fiat}
    \end{subfigure}
    \caption{Bill of Exchange and Money Issuing}
    \label{fig:money}
\end{figure}

Figure~\ref{fig:boe_vor} 
illustrates the macro bookings of a BoE based purchase.
The BoE shifts payments into the future while the product is delivered immediately. 
The are four micro accountings: Seller, Buyer, Seller Bank, and Buyer Bank
and five macro bookings of two micro bookings each:
1: delivery of product: 1a: Seller sells product to Buyer, 1b: Buyer purchases product from Seller.
2: creation of BoE: 2a: Seller issues BoE to Buyer, 2b: Buyer accepts BoE from Seller.
3: "money" exchange for BoE: 3a: Seller sells BoE to Seller Bank, 3b: Seller Bank buys BoE from Seller.
4: transfer of BoE for liabilities: 4a: Seller Bank sells BoE to Buyer Bank, 4b: Buyer Bank buys BoE from Seller Bank.
5: BoE settlement: 5a: Buyer Bank submits BoE to Buyer, 5b: Buyer settles BoE by payment.

The debt flow in the BoE process involves 4 steps.
First internal receivables and liabilities are created at product delivery.
The BoE makes them exchangeable by being a negotiable financial instrument.
A: debt creation (blue arrow - Buyer to Seller):
The Seller delivers the product to the Buyer (1a), creating a receivable for the Seller and a liability for the Buyer (1b).
The Buyer formalises this debt by accepting a BoE issued by the Seller (2a, 2b).
B: BoE monetization (green arrow - Seller to Seller Bank):
The Seller converts the BoE into liquidity by selling it to Seller Bank (3a), receiving a credit in return.
The Seller Bank now holds the BoE as a receivable (3b).
C: BoE transfer (gray arrow - Seller Bank to Buyer Bank):
The Seller Bank transfers the BoE to the Buyer Bank (4a), receiving an equivalent receivable from the Buyer Bank.
The Buyer Bank takes on the liability (4b), holding the BoE as an obligation from the Buyer.
D: final settlement (yellow arrow - Buyer to Buyer Bank):
The Buyer Bank presents the BoE to the Buyer for payment (5a), debiting Buyer’s account.
The Buyer settles its debt (5b), completing the cycle.
Each step transforms the debt, moving it through different financial entities. 
The BoE acts as a transferable financial instrument, facilitating liquidity while maintaining the underlying obligation until final settlement. 
The process creates a dynamic vortex of obligations, rather than closed loop over papers of money.

In the BoE booking, no money is exchanged only deposits or "money" between the agents involved. 
Before money transfers occur, banks usually clear their mutual receivables and liabilities via central bank deposits, possibly followed by settlement in paper money.
Figure~\ref{fig:mon_vor} later illustrates that money primarily exists to facilitate debt repayment on the macro economic level as well.

A BoE can also act as a substitute for money, as a simplified Sola (short of German \emph{Solawechsel}).
A BoE is issued by the drawer and sent to the drawee for acceptance before returning to the drawer.  
A Sola skips this back-and-forth: the drawee creates, signs and sends it to the payee.  

There are two types of Solas (BoEs):
With endorsement: The payee can transfer the Sola to others, e.g. suppliers, by signing on the endorsement list. 
Each endorser shares the liability: if the BoE is unpaid at maturity, the holder can claim repayment from anyone on the list, with recourse up the chain.
Here the restitution and principle of abstraction come in handy.
2. Without endorsement: Acts like anonymous payment; only the drawee is liable to a payee presenting the BoE at maturity.

The endorsement list is like a \emph{blockchain of the renaissance} (or, \emph{a blockchain is like an endorsement list}), enhancing credibility as more parties sign on.
This financial instrument implemented distributed liabilities while facilitating secure, transferable payments, at least since the renaissance.
\subsection{Schumpeter Entrepreneur}\label{sec:kruschum}
The basic building block of a MoMa systems is an investment, defined by Kruschwitz as a time series: 
an initial payout from investor to producer followed by an expected repayment scheme from producer to investor.  
The legal principle of separation informs us that, unlike in \emph{Money is Memory}~\cite{kocherlakota1998money},
memory is given by loan contracts and repayment histories.

\begin{figure}[h!]
    \centering
    \includegraphics[width=1.0\textwidth]{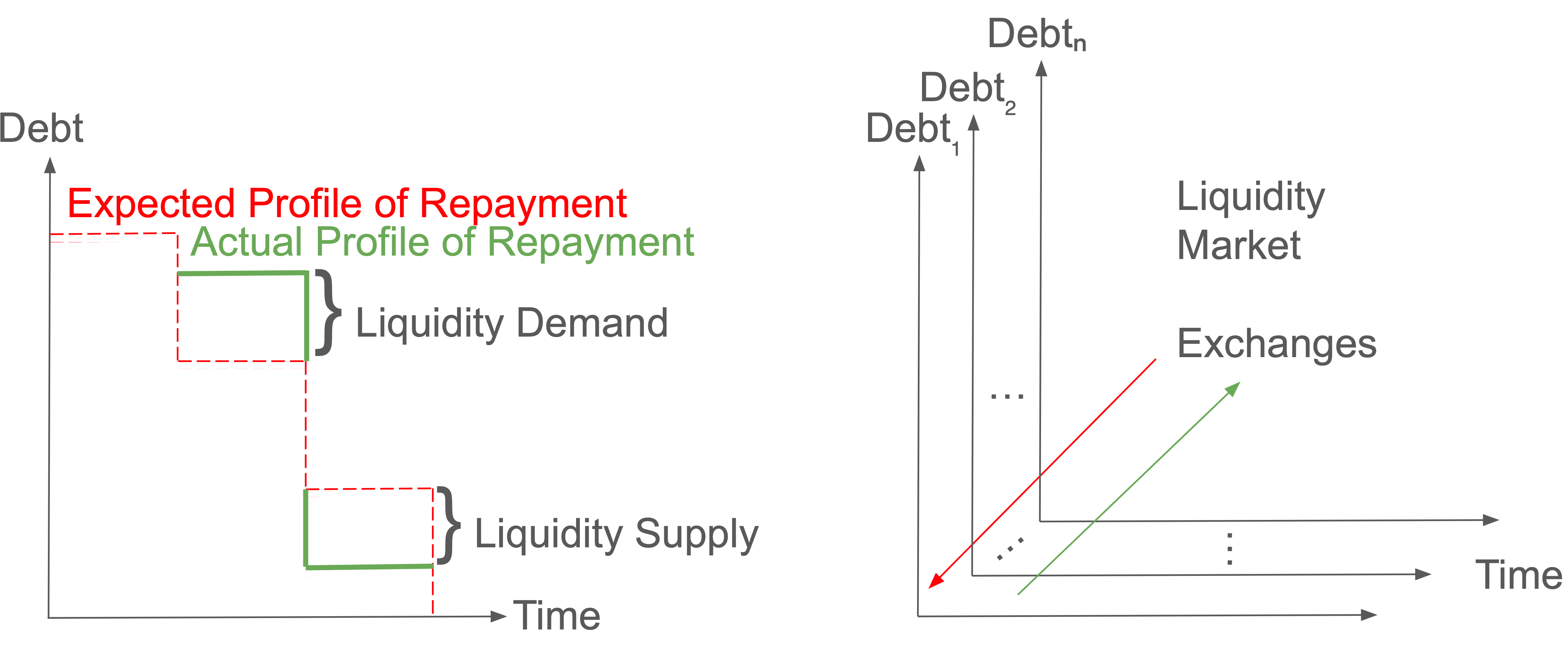} 
    \caption{Kruschwitz Investment of an Schumpeter Entrepreneur}
    \label{fig:kruschwitz}
\end{figure}

Figure~\ref{fig:kruschwitz}
illustrates the expected repayments as a dotted line.
The sales success of producers determines whether actual and expected repayments match.
A saying attributed to Schumpeter is: \emph{Entrepreneurs ride on debt to risky success.}
Deviations of actual from expected repayments lead to Walrasian liquidity markets matching supply and demand, see Figure~\ref{fig:kruschwitz}.
MoMaT answers Hellwig's doubt on Walrasian money markets, by money is traded but not created at markets, which central banks freely do.
\subsection{Banking}\label{sec:banking}
In fiat systems, banks cannot issue the means of payment, like central banks couldn't freely create gold in the gold-based systems. 
Since gold discoveries are random or rather unresponsive to monetary policy, alternative claims like BoEs and bank deposits emerged.  
We apply \emph{economic archaeology} to derive applicable monetary theory, 
in that we study past monetary systems, abstract their functions, and instantiate them in modern technologies for understanding and improving institutional implementations.

The role of banks is illustrated in the bookings of the BoE in Figure~\ref{fig:boe_vor}, showcasing the simplest form of investment in supply chain finance. 
When the Seller Bank purchases a BoE, it invests and bears a risk, if the buyer defaults, the BoE goes to \emph{protest} at a court.
The risk persists until the obligation contract is fulfilled, with the Buyer settling the BoE. 
During this period, the Seller Bank shares joint liability with other endorsers and the drawee. 
In case of default, restitution claims disentangle the losses among the involved agents.  
A bank’s total portfolio of risky investments must balance returns and losses. 
If it fails to repay its debts to the central bank, it becomes illiquid, leading to bankruptcy. 
The interest rate functions as an insurance premium, ensuring that losses are covered before profit is declared.

A fundamental question in monetary theory is whether claims and payments across the hierarchical levels of micro, meso, and macro agents 
align with macro invariances of debts and payments.  
The macro invariance of debt relations 	states that the sum of receivables 	equals the sum of liabilities. 
The macro invariance of payments 		states that the sum of expenditures 	equals the sum of revenues. 
These two invariances are to be consistent as well, as debt invariance pertains to state variables, 
whereas payment invariance concerns flow variables or changes of state variables.  
These invariances are not empirical or behavioural hypothesis
but theoretical necessities for practical, consistent, distributed accounting systems.
A bank has to assure that clearing and settlement possibilities and probabilities between financial BoE or Solas are sufficiently securitised by credibility loans 
by gold, central bank money or other forms of assets.

At the micro level, production generates claims and debts or receivables and liabilities, which are exchanged, cleared, and settled using money or other claims.
At the meso level, banks create receivables and liabilities, clearing and settling them either through the central bank or money payments.  
At the macro level, central banks issue money and deposits for bank. 
They may also refinance or invest in BoEs from companies. 
Deposits at the central bank serve as commands for money, cleared by banks against each other.  
Money issuing is recorded in the central bank’s micro accounting, 
but its role as a means of payment for obligation contracts emerges in macro accounting.
Central banks alone, by monopolistic authority, define and issue money.
\subsection{Supply Chain Finance}\label{sec:scf}
MoMaT suggests money functions to be seen in rather pure form in Supply Chain Finance (SCF), like in Figure~\ref{fig:liquidity}.
A digital BoE based service, such as Dynamic Discounting (DD) on an SCF platform,\footnote{This 
	section benefited greatly from discussions with Markus Wohlgeschaffen, product manager at Traxpay, a pioneer in applications of digital BoEs for SCF.} 
streamlines these processes, making them more efficient and flexible.
In DD, GY (a supplier) has a button in its ERP system, next to invoices issued to VW (the buyer). 
When GY chooses early payment, an accountant presses the button, triggering a transfer of "money" from VW’s to GY’s bank account. 
The digital workflow encompasses four steps:
1. send GY’s request to a digital depository to create a BoE.  
2. A bank purchases the BoE, transferring the dynamically discounted amount to VW to pay GY.
3. At maturity, the bank requests full repayment from VW, earning the discount as a return on its loan.  
4. The BoE is deleted in the digital depository, as the obligation contract is fulfilled by the "money" transfer of the disposition contract without any more needs for money.

The opportunity for services with digital documents, including BoEs, 
arose in UK in 2023 adopting the Model Law of Electronic Transferable Records (MLETR) 
of the United Nations Commission on International Trade Law (UNCITRAL) 
by the Electronic Trade Documents Act (ETDA).\footnote{
  See the Bills of Exchange Act of 1882~\cite{BoEA_1882},
  the German Wechsel Law of 1933~\cite{WG_1933},
  the MLETR of 2017 of UNCITRAL~\cite{mletr_2017}
  and its UK adaptation ETDA~\cite{etda_2023}.
  }
\section{Macroeconomics}\label{sec:macro}
Marx’s critique of capitalism collapses in MoMaT: \emph{money is not created from money}. 
Likewise, the misconception of \emph{interests on interests} fades when interest rates are seen as insurance premiums for cross investment risk.  

The misunderstandings arise from a focus on money circulation rather than on the dynamics of debt (receivables and liabilities) 
forming vortices, not closed loops. 
A banknote changing hands may suggest circulation’s primacy, but legally and economically, debt evolution matters, not the movement of money. 
Money is a means of payment, not the debt itself.  

Geometrically, a vortex spanning time may appear circular when projected along the timeline, 
much like how money’s movement seems cyclical when debt vortices are mapped onto transactions. 
Just as Copernicus replaced Ptolemy’s geocentric model, monetary theory must shift focus: debt vortices, not money flows, define the system. 
To understand this, we examine money issuing under fiat and gold-based systems.
\subsection{Central Banking}\label{sec:cb}

Before the rise of central banks in the 17th century, banks issued banknotes backed by gold, or other valuable resources, making them debt instruments,
a promise by the bank to redeem them for gold.  
This is modelled as credit (or credibility) loans as a financial instrument that extends the BoE into a Sola and two more future contracts on gold and products.
This construction certify banks for proper clearing and settlement size and timing.

Founded in 1609, the Wisselbank of Amsterdam~\cite{wissel_bank} (\emph{Wissel} is the Dutch word for the German \emph{Wechsel} or BoE) pioneered international payments, crucial for European colonial trade, which required investments in shipping. 
Long voyages, high risks, and delayed revenues made financing essential. 
From this model, central banks emerged, including the Bank of England, which managed national currencies like the Pound, backed by Sterling silver. 
Early central banks operated on gold or silver debt, issuing banknotes as documented obligations.

Figure~\ref{fig:gold} shows banknotes issuing in a gold-based system.
The bookings involve two steps.
Step 1: issuing a banknote (1a):
The central bank prepares to buy gold, effectively expanding the money supply by recording a warehouse receipt.
Step 2: gold transaction (2a, 2b):
The central bank purchases gold from a gold miner (2a).
The gold miner receives a banknote in exchange for gold (2b), serving as a claim on gold.
Booking 3a records the issuance of a Sola, following the same process as issuing a banknote (1a).

Gold was the legal means of payment, but to avoid transferring it in every transaction, banknotes were introduced as claims on gold, redeemable at the central bank. Essentially, a gold miner deposited gold in the central bank, receiving a banknote in return. 
This note circulated as a transferable payment instrument within a purchase obligation, allowing any holder to exchange it for gold.

In 1971, Nixon ended the gold-based systems, making paper money the basis of contracts under the fiat system. 
Unlike gold-based banknotes, paper money does not represent central bank debt and lacks inherent claims, 
meaning \emph{Issued Paper Money} records circulation, not obligations. 
Without a claim, there is no corresponding liability.

The bookings of issuing paper money in the fiat system follows 5 steps.
1: money issuing (1a):
The central bank (CB) issues paper money, legally registering it in the \emph{Issued Paper Money} database.
2: Loan to a Bank (2a, 2b):
The CB grants a loan to a commercial bank.
2a: The CB records an increase in receivables and liabilities.
2b: The bank records an increase in its credit and liabilities.
3: Bank withdraws money from CB (3a, 3b):
The bank transforms its option right into a due claim, thereby ordering the central bank to hand over the requested amount of paper money.
3a: The CB reduces its liabilities and cash holdings.
3b: The bank decreases its central bank deposit but increases its cash holdings.
4: Bank grants a loan to a company (4a, 4b):
The bank extends a loan to a company, creating new "money", i.e. new deposits.
4a: The bank records an increase in receivables and liabilities.
4b: The company records an increase in credits and liabilities.
5: Company withdraws cash (5a, 5b):
The company converts part of its bank deposit into "money" for payments (e.g. wages).
5a: The bank reduces its liabilities and cash reserves.
5b: The company decreases its bank deposit but increases its cash holdings.

Issuing a Sola (6a) follows the same principle as creating paper money (1a). 
Unlike gold-based money, paper money exists by legal decree, allowing unlimited issuance. 
It enters circulation through loans and withdrawals, enabling transactions without commodity backing.
The Sola in 6a illustrates that issuing paper money can be seen as similar to a Sola. 
However, unlike a traditional Sola or a warehouse receipt, which represents a claim (e.g. gold in the gold-based system), fiat money has no underlying asset to claim. 
In the gold-based systems, central bank liabilities were backed by gold, effectively balancing debt. 
Gold itself is neither a claim nor a liability like paper money, which carries no inherent claim. 
Fiat systems require no backing, only legal declaration as a means of payment.
In a fiat system, the central bank starts without debt, just as gold and banknotes balanced out in the gold-based systems. 
Money settles debt but is not debt itself. 
Since Nixon ended the gold-based system, central banks cannot go bankrupt in their own currency, 
making fiat systems bankrun resistant as an accidental feature, not a flaw. 
No gold or intrinsic value for paper money is needed for the system to function.

Money issuing should align with productive economic activity and investment driven debt needs. 
MoMaT identifies SCF variables being key for monetary policy, now easily tracked via receivables and liabilities in company ERP and accounting systems. 
Monetary policy and liquidity platforms must consider liquidity demand and supply. 
Until the 1990s, the Bundesbank used BoE refinancing and the discount rate as policy tools, accompanied by interbank market interest rates and
direct interventions at asset markets.

MoMaT defines central bank policy involving three topics: 
1: injecting money into the economy through bank loans, creating central bank deposits to create solvency,
2: setting a minimum credibility rating for such loans, 
and 3: managing the payment system and macro accounting with financial regulations. 
Following Bagehot’s principle, MoMaT advocates unlimited crisis lending at high costs while leaving inflation to government income policies. 
Unlike conventional monetary economics, MoMaT sees no need for artificial scarcity, money supply should allow to match the volume of unsettled receivables and liabilities.

In the gold-based system, central banks acted as lenders of last resort, but this failed when they ran out of gold, leading Nixon to abandon the system. 
In the fiat system, central banks serve as first liquidity providers, able to issue money infinitely. 
When black swan events and banks fail, central banks can insure losses through bailouts, effectively smoothing investment losses as consumption costs. 
These bailouts can be distributed at the highest societal level, with the central bank as the ultimate risk sharing entity.
\subsection{Monetary Vortex}\label{sec:vortex}
\begin{figure}[ph!]
    \centering
    \includegraphics[angle=90,width=0.64\textwidth]{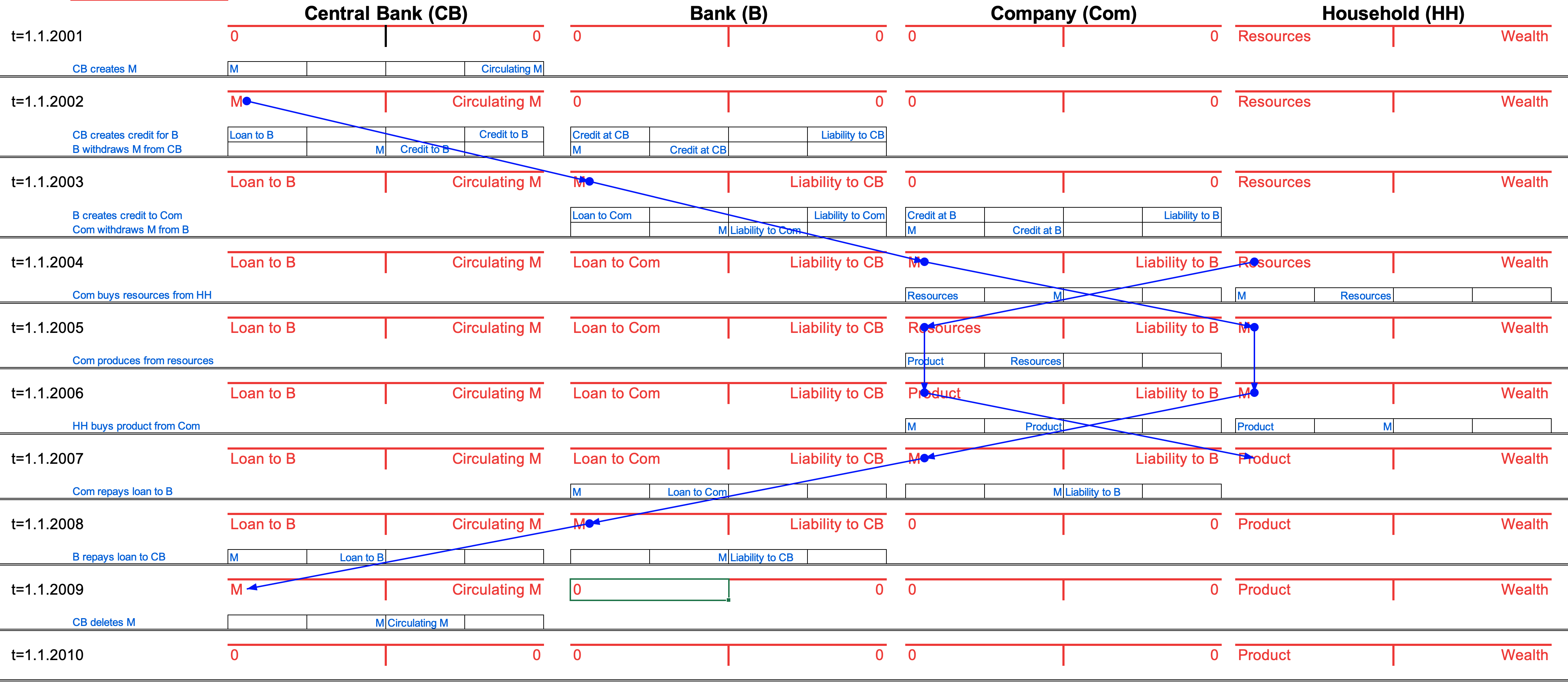} 
    \caption{Monetary Vortex}
    \label{fig:mon_vor}
\end{figure}
In Figure~\ref{fig:mon_vor}, 
we illustrate that vortices of debt, rather than circulating money, are at the core of the dynamics of monetary systems.  
The macro accounting framework includes the Central Bank, Bank, Company, and Household sectors, where resources, products, liabilities, and money are created, exchanged, and deleted throughout the economy’s productive cycles.

At t = 1.1.2001\footnote{Balance 
	sheets report state variables at a specific time, e.g. at t = 1.1.2001. 
	Bookings in financial statement record flow variables as changes of state variables over a period, e.g. in t=2001.
	}, the Household sector 
holds resources with no liabilities, making assets equal equity (E = A). 
The Central Bank issues money as a sovereign act, recorded solely in its own micro accounting system
like alchemically creating gold in a gold-based system, but without intrinsic value or backing. 
It is the only singleton (micro accounting) booking in a macro accounting system.
The central bank’s balance sheet records cash (M) as an asset and \emph{Circulating M} as a liability, marking registered paper money.
The central bank then grants loans as credits.  
t = 2002: The central bank lends to the bank.  
t = 2003: The bank withdraws cash and lends to a company.  
t = 2004: The company buys resources or work from households.  
t = 2005: The company produces goods.  
t = 2006: The household buys the product, returning cash to the company.  
t = 2007: The company repays its loan to the bank.  
t = 2008: The bank repays the central bank.  
t = 2009: The central bank destroys the money.  
t = 2010: Balance sheets reset to zero.  

This process transforms resources into products or GDP, illustrating an economy characterised by a division of labor
where monetary vortices and debt repayments drive production and exchange over time in monetary macro accounting (MoMa).
The monetary vortex in Figure~\ref{fig:mon_vor} charts the flow of money (M) through balance sheets, from issuing to deletion at the central bank.  
Conversely, the physical vortex traces the conversion of resources into products, moving counter to the flow of money.
When viewed along the time axis, top to bottom, the interplay of debts, money, resources, and products creates the illusion of circular flows, 
much like in Figure~\ref{fig:liquidity}.
Money and debt bridge the time gap in production, transforming resources into products and enabling exchange of the GDP. 
While this may appear as a circular flow, it is a timed vortex of resources and money. 
Once all debts are repaid (t = 1.1.2010), money can be deleted. 
Reusing the same money in new debt cycles only creates circulation of money, new and closing parallel debt relations.
At t = 1.1.2010 wealth and GDP distribution in the household sector reflects how investment risks and failures were absorbed by banks and how resources, labor, and investments were financed from t = 2002 to t = 2009. 

Money functions not just to coordinate labor but also to allocate risk and distribute economic output, the GDP. 
Modeling risk requires stochastic processes to account for investment failures, treating losses as consumption or costs. 
Failed investments should not be hidden in \emph{bad banks} or bailed out without a theoretical and political framework.
The dynamics of debts and money and resources and products of the GDP as vortices 
is akin to our solar system's planets 
\begin{figure}[t!]
    \centering
    \includegraphics[width=0.5\textwidth]{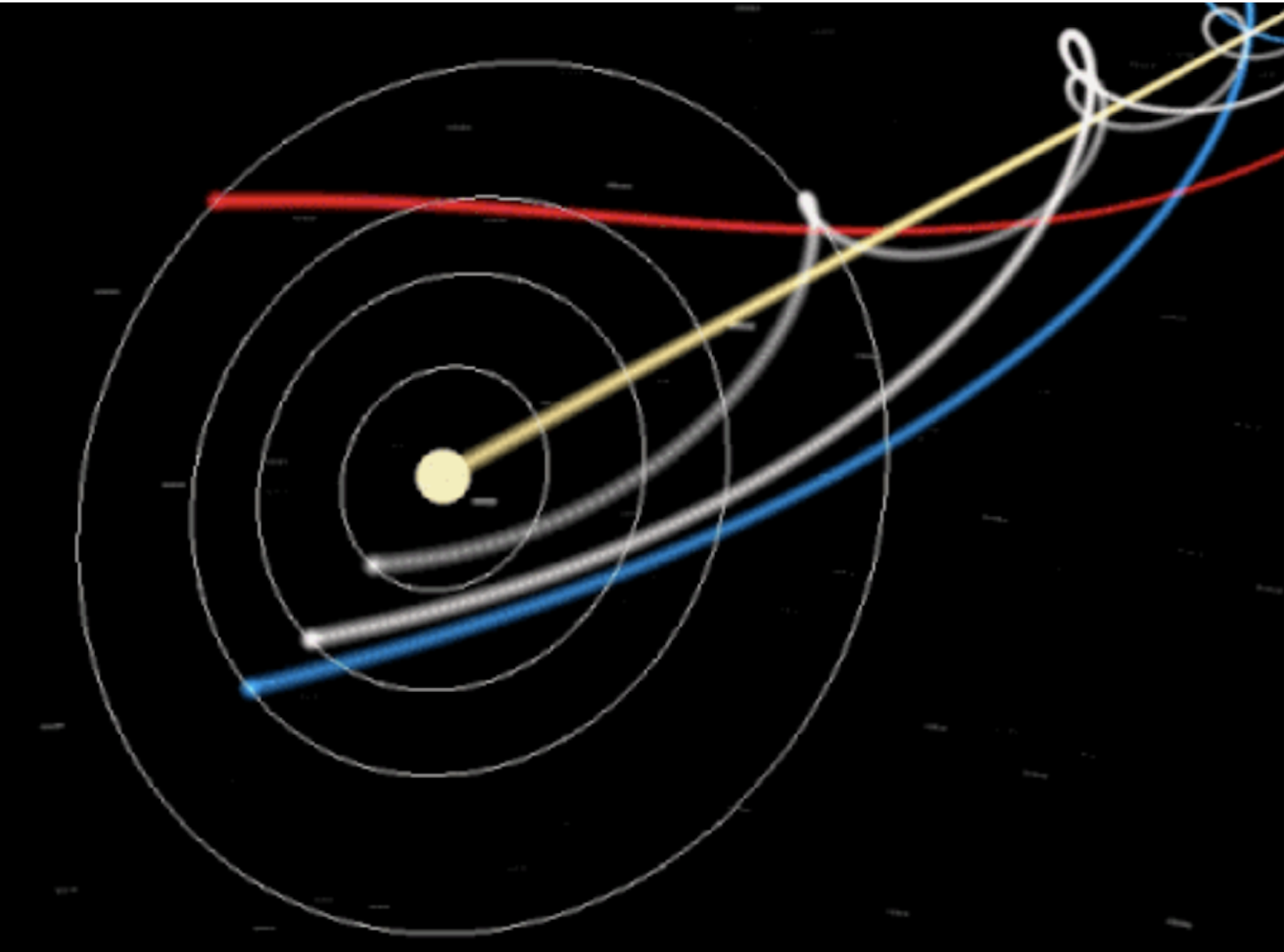} 
    \caption{Planetary Vortices}
    \label{fig:plan_vor}
\end{figure} 
in Figure~\ref{fig:plan_vor},
see the movie at~\cite{solar_vortex}.
The planets do not orbit in circles or ellipses around the sun but in vortices, 
following the sun’s trajectory as it travels at approximately 70,000 km/h around the galactic center.  
This motion appears as a circular orbit only when projected onto a two-dimensional plane through the sun, 
effectively neglecting the space-time direction in which the sun itself is moving.

Paper money circulation masks debt repayments in anonymous transactions.
Vortices can be found as money transactions in endorsement lists of BoEs. 
Anonymity is like projecting planetary orbits and debt relations on a plane through the sun or on papers of money. 
To address monetary theory and policy challenges, we must link receivables, liabilities, expenditures, and revenues 
across economic micro, meso and macro layers with properly typed financial contracts.

We have seen that endorsement lists of BoEs have a similar functionality as blockchains.
These lists are representing chains of debt relationships in which a claim document is passed along.
In this sense, they serve as a model for blockchains as databases of ownership rights to claims which can be maintained by an authorised depository.
In payment processes recorded in endorsement lists or blockchains, we could actually find the monetary vertices discussed above as data ready to be analysed for liquidity supply and exchange purposes or for analysis for the conduct of a monetary policy. 

In this application of an "archeology of economics" the abstract economic functionality using records of debtors and creditors, actually even bilateral solvency checks, 
using the historical technology of pencil and paper is to be transferred and implemented in modern technologies of digital computing and then generalised, 
for example to smart contracts and applications of blockchains or other new technologies.
The German and international laws of the BoE are treasuries of historical experience and specification of financial needs - to be potentially used as a pool
of ideas for smart contracts design in modern times.
\section{Mathematics}\label{sec:math}
The challenge in monetary theory and policy is to model a Monetary Macro Accounting (MoMa) system in Figure~\ref{fig:mon_vor}. 
MoMa models national accounts as networks of debt contracts across banks, companies, and households, using a (reduced form) Markov model based only on past data. Analysing this requires MoMaT with structural econometric models with forward-looking agents, as in~\cite{Winschel2010},~\cite{Heiss2008} for econometrics or \cite{kochenderfer2022algorithms} for multi-agent AI, both programmable with open games.

\begin{figure}[htbp]
    \centering
    \includegraphics[width=0.75\textwidth]{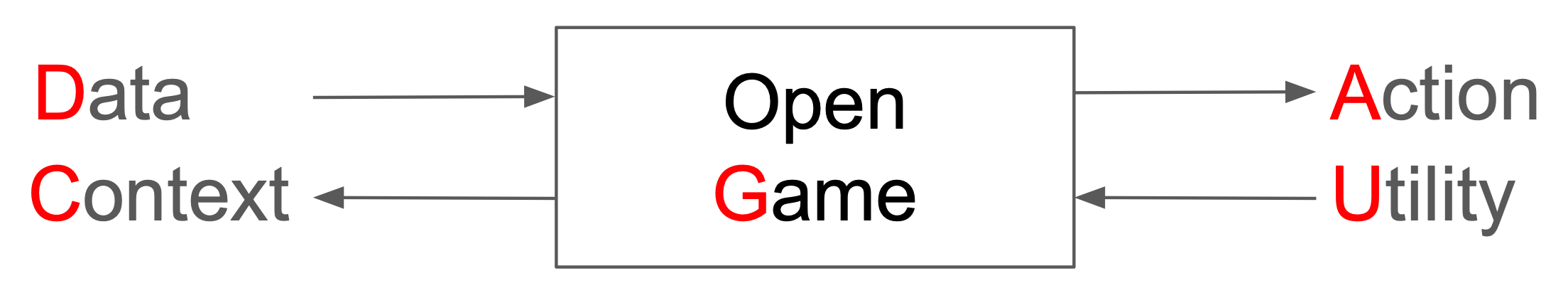}
    \caption{Open Game}
    \label{fig:og}
\end{figure}

An Open Game~\cite{ghani2018compositional} is depicted in Figure~\ref{fig:og} as a string diagram, with a probabilistic extension in~\cite{bolt_bayesian_2023}.
Open games represent time bidirectionally: past to future (left to right) and future to past (right to left). 
Decisions form one- or multi-person games, built from smaller n-person games into larger n+1-person games via sequential and parallel combinators. 
In category theory, these correspond to function composition and tensor product, providing a process-oriented approach to mathematics \cite{dagstuhl_road}. 
Parallel and sequential processes are composed through concatenation and juxtaposition in string diagrams, while hierarchical composition uses equational substitution, 
enabling both white-box and black-box economic modelling.\footnote {
	\emph{White boxing} means \emph{making explicit} by showing the content of a \emph{black box}.
	We can use the \emph{white box} $a= b +  c$ to show in $d =  a +  e$ the \emph{black box} $a$ as $d= a+e=b +  c+ e$.
	}
In a string diagram the box represents functions (or processes or morphism in category theory) 
with the open games type:
\[
G:D\times C^*\rightarrow A\times U^*
\]
Open games are compositional because Data, Action, Context, and Utility are also open games, and their compositions remain open games.
The superscript $^*$ denotes contravariance in teleology, goal, utility, belief, or anticipation where arrows point left, opposite to usual time flow. 
Open games capture the Aristotelian teleological causality, linking the future to the present (purpose). 
The other three causes, normative (plan), material (bricks), and effective (work) define physical existence (e.g. of a house), 
while teleology defines purpose (e.g. living in a house).  

String diagrams are not just representations but formal mathematical objects with semantics for proofs. 
Open games provide a visual programming language for forward-looking economic models, making them executable software with provable properties. 
As both models and runnable programs, they leverage computer science and mathematics to understand and implement economies.

Since open games are similar to the interdisciplinary research on quantum computing programming languages,
economic theories formulated in open games are quantum ready.\footnote{
	The papers on coalgebraic games~\cite{Abramsky2012a},~\cite{dusko2009} are precursors to open games~\cite{ghani2018compositional}
	inspired by the work~\cite{AbramskyCoecke2004} on quantum protocols.
	}
Therefore transactions of MoMa systems with open games as smart contracts can be about quantum secured assignments and transfers of property rights with non-deletable data with free write and read access on blockchains or distributed ledgers.

Open games represent bidirectional computation, including time flow, relations, double-entry accounting, constraint propagation, or smart contracts. 
They can model structural econometrics, multi-agent systems, belief formation, or strategic decisions. 
With built-in back propagation \cite{fong2019backprop}, they also support parameter estimation and data science algorithms.

Macro invariances can be modelled using sheaf theory over micro accounting automata \cite{rambaud2010algebraic}, embedded in a network of contracts. 
A compositional semantics for MoMa must derive macro behaviour from micro accounting systems interconnected at the meso level. 
Constraints from micro, meso and macro invariances should be transparently booked. 
State transitions follow initial algebras in gold-based systems, 
while fiat systems can use final coalgebras to model infinite, behavioural structures for booking smart contracts in macro accounting.

Category theory reorganises mathematics~\cite{dagstuhl_road} as its compositional theory.
It can provide semantics to economics in coalgebras~\cite{dagstuhl_econ}.\footnote{The discovery of open game structures 
	emerged from the search for a mathematical framework addressing Lucas' critique and reflexivity in social sciences~\cite{reflexivity_wiki}. 
	Reflexive structures in mathematics, such as lambda calculus, higher order functions, 
	Turing machines, and formal semantics in theoretical computer science \cite{pavlovic_programs_2023}, provide a foundation. 
	Meta circular interpreters, like those by Graham~\cite{roots_of_lisp_2001} and McCarthy~\cite{mccarthy1960recursive}, illustrate this with Lisp implementing itself.
	Social sciences and economics offer complex epistemological situations
	where modern semantic tools like category theory, type theory, or coalgebras can enhance institutional modelling and implementations.
	}
A compositional theory of monetary macro accounting 
needs to zoom in and out~\cite{willems_behavioral_2007} at different behavioural levels of the systems.
Systems are to be composed and decomposed while keeping the levels looked at in a consistent state.
Global and local consistencies and inconsistencies of behaviours interact
and we need to understand how local goals and obstacles can be resolved into a consistently behaving global whole.

Sheaf theory constructs complex models by combining local information into a consistent whole \cite{robinson_sheaf_2016, schenck_algebraic_2023}. 
Like two eyes merging fields of view or an atlas forming a globe, sheaves splice local data into a global structure. 
In equations, local solutions integrate into a global solution, just as local function approximations combine into a globally defined function. 
Similarly, Bayesian networks update through local observations, propagating until the entire global model aligns with the data.

A MoMa system integrates economic agents are resource owners, consumers, producers, banks, and central banks
each maintaining local micro accounting of debt and payments. 
These agents form a network where nodes represent entities and edges track financial flows, ensuring consistency 
through receivables, liabilities, expenditures and revenues.
Aggregated into macro accounting, this network maintains stability via macro invariances. 
In sheaf theory, the \emph{global section} represents systemic consistency, which monetary policy must uphold. 
During crises, its role is to restore the system to this stable state.

\begin{figure}[htbp]
    \centering
    \includegraphics[width=0.25\textwidth]{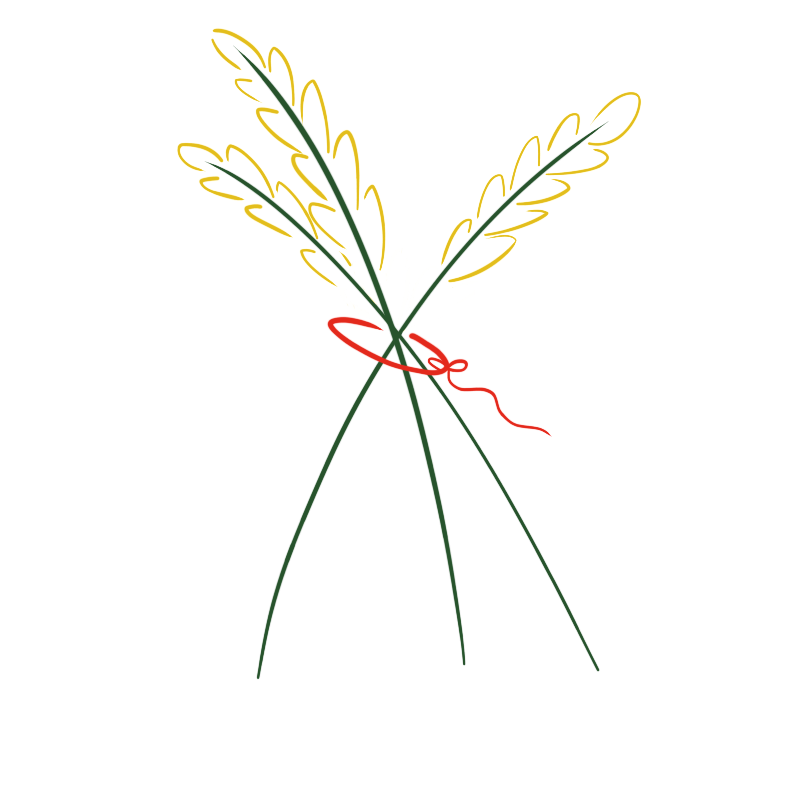}
    \caption{Agricultural Sheaf}
    \label{fig:sheaf}
\end{figure}

The name \emph{sheaf} for these mathematical structures is an analogy to agricultural sheaves in Figure~\ref{fig:sheaf}.
Stalks represent micro accounting data atop the network of local agents, while the macro accounting as a sheaf forms the global structure, 
bound by consistency conditions, analogous to the thread in the agricultural sheaf.
Stability emerges in the sheaf, a property individual stalks alone do not have, 
much like money’s functionality emerges only at the macro accounting level from booking its issuing in the central bank's micro accounting.

Homology theory identifies obstacles to global consistency and macro invariances, particularly when inconsistencies arise in specific regions or points in time. 
In MoMaT, these may appear as local liquidity imbalances or bad banks, resolved by issuing BoEs or some monetary policy across micro, meso, and macro levels. 
This enables a decentralised, automated monetary policy, reducing reliance on discretion. 
Automated policy is possible, debated as \emph{Taylor functions}, which optimally set central banks' interest rates based on inflation or the phase in the business cycles.
\section{Conclusion}\label{sec:conc}
\emph{What is money?} is answered by
money being the thing that settles obligation and debt relations and fulfils them by disposition of property rights over money and products.
Money as a technology within macro accounting enables division of labor, risk allocation, and GDP distribution across micro, meso, and macro levels of an economy.  

At the micro level, money bridges the time gap in production, allowing producers to pay suppliers before receiving payment from customers. To do this, producers take on debt, repaying it with future earnings. Legally, money fulfils obligations in contracts
either as payment for goods or repayment of loans without needing intrinsic value. 
Monetary units are arbitrary but serve as the legal means to settle debts.  
At the meso level, banks manage investment risks by pooling loans and balancing returns through interest rates, functioning as insurance premiums. 
They provide solvency for companies by guaranteeing this solvency at their own risk.
At the macro level, central banks issue money and release it to banks for investment. 
They absorb non-insurable risks as lenders of first liquidity in fiat systems and lenders of last resort in gold-based systems.  

Sheaves can structure networks of debts and repayments, ensuring consistency across economic levels through macro accounting invariances. 
Homology theory can correct local inconsistencies, supporting automated monetary policy. 
Open games can serve as modelling and computational tools for deductive types in MoMa systems
and inductive types in MoMaT for economic forecasts, back propagation, econometric, and multi-agent AI analysis.  

MoMaT defines the functionality of money as a means of payment
and the unit of account for the settlement of monetary obligations 
in two-sided contracts governing obligations and dispositions of property rights.
The store of value functionality can be modelled as financial BoEs or Solas in credit (credibility) loans.
The means of exchange functionality of money provides a numéraire, and medium of exchange.

Finally, we have provided a link to the blockchain world. 
The endorsement lists of the BoEs are designed to record exchanges of the claims to the payments at maturity date.
Similarly, blockchains are linked chains of exchanges of claims. 
We hope that the efforts in blockchain technologies to provide a permissionless and non-deletable record keeping technology for the transfers of property rights
can take advantage of our monetary theory for the goal to provide a new approach to monetary systems.
\bibliographystyle{plainnat}
\bibliography{bib_items}
\end{document}